\documentstyle[12pt]{article}

\catcode`\@=11
\long\def\@makefntext#1{
\protect\noindent \hbox to 3.2pt {\hskip-.9pt
$^{{\ninerm\@thefnmark}}$\hfil}#1\hfill}		

\def\@makefnmark{\hbox to 0pt{$^{\@thefnmark}$\hss}}  

\def\ps@myheadings{\let\@mkboth\@gobbletwo
\def\@oddhead{\hbox{}
\rightmark\hfil\ninerm\thepage}
\def\@oddfoot{}\def\@evenhead{\ninerm\thepage\hfil
\leftmark\hbox{}}\def\@evenfoot{}
\def\sectionmark##1{}\def\subsectionmark##1{}}

\renewcommand{\thefootnote}{\fnsymbol{footnote}}

\newcounter{sectionc}\newcounter{subsectionc}\newcounter{subsubsectionc}
\renewcommand{\section}[1] {\vspace*{0.6cm}\addtocounter{sectionc}{1}
\setcounter{subsectionc}{0}\setcounter{subsubsectionc}{0}\noindent
	{\normalsize\bf\thesectionc. #1}\par\vspace*{0.4cm}}
\renewcommand{\subsection}[1] {\vspace*{0.6cm}\addtocounter{subsectionc}{1}
	\setcounter{subsubsectionc}{0}\noindent
	{\normalsize\it\thesectionc.\thesubsectionc. #1}\par\vspace*{0.4cm}}
\renewcommand{\subsubsection}[1]
{\vspace*{0.6cm}\addtocounter{subsubsectionc}{1}
	\noindent {\normalsize\rm\thesectionc.\thesubsectionc.\thesubsubsectionc.
	#1}\par\vspace*{0.4cm}}

\newcounter{appendixc}
\newcounter{subappendixc}[appendixc]
\newcounter{subsubappendixc}[subappendixc]

\renewcommand{\appendix}[1] {\vspace*{0.6cm}
        \refstepcounter{appendixc}
        \setcounter{figure}{0}
        \setcounter{table}{0}
        \setcounter{equation}{0}
        \renewcommand{\thefigure}{\Alph{appendixc}.\arabic{figure}}
        \renewcommand{\thetable}{\Alph{appendixc}.\arabic{table}}
        \renewcommand{\theappendixc}{\Alph{appendixc}}
        \renewcommand{\theequation}{\Alph{appendixc}.\arabic{equation}}
        \noindent{\bf Appendix \theappendixc #1}\par\vspace*{0.4cm}}

\def\abstracts#1{{

\centering{\begin{minipage}{12.2truecm}\footnotesize\baselineskip=12pt\noindent
	\centerline{\footnotesize ABSTRACT}\vspace*{0.3cm}
	\parindent=0pt #1
	\end{minipage}}\par}}

\renewenvironment{thebibliography}[1]
	{\begin{list}{\arabic{enumi}.}
	{\usecounter{enumi}\setlength{\parsep}{0pt}
\setlength{\leftmargin 1.25cm}{\rightmargin 0pt}
	 \setlength{\itemsep}{0pt} \settowidth
	{\labelwidth}{#1.}\sloppy}}{\end{list}}

\newcounter{itemlistc}
\newcounter{romanlistc}
\newcounter{alphlistc}
\newcounter{arabiclistc}

\newcommand{\fcaption}[1]{
        \refstepcounter{figure}
        \setbox\@tempboxa = \hbox{\footnotesize Fig.~\thefigure. #1}
        \ifdim \wd\@tempboxa > 6in
           {\begin{center}
        \parbox{6in}{\footnotesize\baselineskip=12pt Fig.~\thefigure. #1}
            \end{center}}
        \else
             {\begin{center}
             {\footnotesize Fig.~\thefigure. #1}
              \end{center}}
        \fi}

\newcommand{\tcaption}[1]{
        \refstepcounter{table}
        \setbox\@tempboxa = \hbox{\footnotesize Table~\thetable. #1}
        \ifdim \wd\@tempboxa > 6in
           {\begin{center}
        \parbox{6in}{\footnotesize\baselineskip=12pt Table~\thetable. #1}
            \end{center}}
        \else
             {\begin{center}
             {\footnotesize Table~\thetable. #1}
              \end{center}}
        \fi}

\def\@citex[#1]#2{\if@filesw\immediate\write\@auxout
	{\string\citation{#2}}\fi
\def\@citea{}\@cite{\@for\@citeb:=#2\do
	{\@citea\def\@citea{,}\@ifundefined
	{b@\@citeb}{{\bf ?}\@warning
	{Citation `\@citeb' on page \thepage \space undefined}}
	{\csname b@\@citeb\endcsname}}}{#1}}

\newif\if@cghi
\def\cite{\@cghitrue\@ifnextchar [{\@tempswatrue
	\@citex}{\@tempswafalse\@citex[]}}
\def\citelow{\@cghifalse\@ifnextchar [{\@tempswatrue
	\@citex}{\@tempswafalse\@citex[]}}
\def\@cite#1#2{{$\null^{#1}$\if@tempswa\typeout
	{IJCGA warning: optional citation argument
	ignored: `#2'} \fi}}

 1
 1
 1

\font\ninerm=cmr9

\newcommand{\be}{\begin{equation}}
\newcommand{\ee}{\end{equation}}
\newcommand{\half}{\frac{1}{2}}
\begin{document}

\centerline{\normalsize\bf
Mass Bounds in Mirror-Fermion Models}

\centerline{\footnotesize LEE LIN}
\baselineskip=13pt
\centerline{\footnotesize\it Department of Physics}
\baselineskip=13pt
\centerline{\footnotesize\it National Chung Hsing University}
\baselineskip=12pt
\centerline{\footnotesize\it Taichung 40227, Taiwan, ROC}
\centerline{\footnotesize E-mail: llin@phys.nchu.edu.tw}

\vspace*{0.9cm}
\abstracts{
Numerical simulations are performed on different lattice sizes
of chiral U(1) and SU(2) scalar-fermion models with explicit
mirror pairs of fermions in the broken symmetry phase. Relevance of
these models to the electroweak theory is discussed.
Shift symmetry of the action is exploited to decouple the unwanted
mirror-fermion. Upper and lower bounds on the renormalized scalar
mass as a function of the renormalized fermion mass are obtained
in the SU(2) version of the model.
Our numerical data are found to be in qualitative agreement with
one-loop results. No evidence for a nontrivial fixed point has
been observed. The mirror-fermion models are likely to be trivial
in the continuum limit. Our Monte Carlo data show that with a $180\, GeV$
top quark, the Standard Model Higgs boson should have a mass between
$100$ and $800\, GeV$.}

\normalsize\baselineskip=15pt
\setcounter{footnote}{0}
\renewcommand{\thefootnote}{\alph{footnote}}

\section{Introduction}

It is believed that the only possible nonperturbative (i.e. strongly
interacting) sectors in the electroweak theory of the minimal Standard
Model are the Higgs and heavy fermion sectors. (Here heavy or light is
meant to be compared to $100\, GeV$ scale, the energy scale of the weak
interaction.) If the electroweak theory is trivial in the infinite cutoff
limit, then we can obtain upper and lower bounds on the Higgs mass as a
function of the heavy fermion mass.
Since those sectors can become strongly interacting, it is important that
nonperturbative investigation be carried out. In this report, the
nonperturbative method is Monte Carlo simulation on the lattice.

Because there are lots of bare parameters in the electroweak theory, we
need to make some approximations before lattice calculations become
feasible. The approximation we make is to
ignore all the gauge couplings assuming that all gauge
couplings produce perturbative effects. (Although QED has a Landau pole,
its coupling at $100\, GeV$ scale can still be safely ignored because
Landau pole presumably appears at some astronomically high scale.)
We call this the zeroth-order approximation.
We also ignore all the light fermions. Therefore, the only fermion left
is the top quark (and/or the fourth family fermions should they exist).

It is well known that naive fermion action on the lattice suffers from
species doubling. The ways to handle fermion doublers are the Wilson
fermion and the staggered fermion~\cite{WS}.
 In this report, we take the approach
of the Wilson fermion. However, the electroweak theory is a chiral gauge
theory. Even after we ignore the gauge couplings, we still have global
chiral symmetries left. The naive Wilson fermion action breaks the chiral
symmetry explicitly and is not good for the electroweak theory. In order
to cure this, two models were proposed. One is the Smit-Swift
model~\cite{SS}. The other is the mirror-fermion model~\cite{IM}.
We will report on our lattice study of
the electroweak theory using the mirror-fermion model
in the zeroth-order approximation.

The global symmetry of the mirror-fermion model can be U(1) or SU(2).
It is found that the U(1) and SU(2) versions have the same qualitative
properties. Since SU(2) symmetry is closer to the electroweak part of
the real world, we will concentrate on the SU(2) mirror-fermion model.

\section{Lattice Action and Decoupling Limit}

The lattice action of the mirror-fermion model is a sum of the O(4)
($\cong \rm SU(2)_L \otimes SU(2)_R$) symmetric pure scalar part
$S_\varphi$ and fermionic part $S_\Psi$:
\be \label{eq01}
S = S_\varphi + S_\Psi \,.
\ee
$\varphi_x$ is the $2 \otimes 2$ matrix scalar field, and
$\Psi_x \equiv (\psi_x, \chi_x)$ stands for the mirror pair of fermion
doublet fields (usually $\psi$ is the fermion doublet and $\chi$ the
mirror fermion doublet).
In the usual normalization conventions for numerical simulations we
have
$$
S_\varphi = \sum_x \left\{ \half {\rm Tr\,}(\varphi_x^+\varphi_x) +
\lambda \left[ \half{\rm Tr\,}(\varphi_x^+\varphi_x) - 1\right]^2
- \kappa\sum_{\mu=1}^4
{\rm Tr\,}(\varphi^+_{x+\hat{\mu}}\varphi_x)
\right\} \ ,
$$
$$
S_\Psi = \sum_x \left\{ \mu_{\psi\chi} \left[
(\overline{\chi}_x\psi_x) + (\overline{\psi}_x\chi_x) \right]
\right.
$$
$$
- K \sum_{\mu=\pm 1}^{\pm 4} \Bigl[
(\overline{\psi}_{x+\hat{\mu}} \gamma_\mu \psi_x) +
(\overline{\chi}_{x+\hat{\mu}} \gamma_\mu \chi_x)
$$
$$
+ r \left( (\overline{\chi}_{x+\hat{\mu}}\psi_x)
- (\overline{\chi}_x\psi_x)
+ (\overline{\psi}_{x+\hat{\mu}}\chi_x)
- (\overline{\psi}_x\chi_x)  \right) \Bigr]
$$
\be \label{eq02}
\left.
+ G_\psi \left[ (\overline{\psi}_{Rx}\varphi^+_x\psi_{Lx}) +
(\overline{\psi}_{Lx}\varphi_x\psi_{Rx}) \right]
+ G_\chi \left[ (\overline{\chi}_{Rx}\varphi_x\chi_{Lx}) +
(\overline{\chi}_{Lx}\varphi^+_x\chi_{Rx}) \right]
\right\} \,.
\ee
Here the two fermion fields $\psi$ and $\chi$ are Wilson fermions,
$K$ is the fermion hopping parameter, $r$ the Wilson-parameter,
which will be fixed to $r=1$ in the numerical simulations, and the
indices $L,R$ denote, as usual, the chiral components of fermion
fields. One can easily see that the above action
has an explicit chiral symmetry in the sense that left- and
right-handed components of the fermion fields can transform
independently. However, the right-handed $\chi$-field has to
transform exactly as the left-handed $\psi$-field, and left-handed
$\chi$-component as the right-handed $\psi$-component. (So an
off-diagonal bare fermion mass term can exist in the action.)
Therefore, $\psi$ and $\chi$ are mirror partners of each other.
In our simulations, we take normalization that
fermion--mirror-fermion mixing mass is $\mu_{\psi\chi}=1-8rK$.

The fermionic part $S_\Psi$ is given here for a single mirror
pair of fermions.
Taking the adjoint transforms fermions to mirror fermions and vice
versa, but as noted before, without
$\rm SU(3)_{colour} \otimes U(1)_{hypercharge}$ gauge couplings they
are equivalent to each other.

All renormalized quantities can be defined. (See~\cite{LMMW} for
details.) Since the symmetry of the action does not exclude the
mixing fermion mass term, we will have a nonzero renormalized
fermion mixing mass $\mu_R$. The physical fermion states will be linear
combinations of the renormalized $\psi$ and $\chi$ fields in general.
One can always tune the bare parameters to have a zero renormalized
mixing. It turns out that in most cases this tuning is rather
difficult and time consuming in the simulation.

So far we have not discovered mirror-fermions in the
experiments, we do not want to have them in the physical spectrum.
There are different ways to decouple those unwanted fermions.
We can tune the bare parameters such that mirror-fermion has its mass
of the order of the cutoff scale. Hopefully this heavy mirror-fermion
will decouple.
The more elegant way is to take the advantage of the symmetry of
the model and is explained below.

The action shown in eq.(\ref{eq02}) is invariant under the following
transformations of the fermion fields
\be \label{eq03}
\psi_x\,\rightarrow\,\psi_x\,+\,\omega\,\, ,\,\,
\bar\psi_x\,\rightarrow\,\bar\psi_x\,+\,\bar\omega\,\,\,\,\,
{\rm if}\,\,\, G_\psi=0\,\, ;
\ee
\be \label{eq04}
\chi_x\,\rightarrow\,\chi_x\,+\,\epsilon\,\, ,\,\,
\bar\chi_x\,\rightarrow\,\bar\chi_x\,+\,\bar\epsilon\,\,\,\,\,
{\rm if}\,\,\, G_\chi=0\,\,
\ee
where $\omega$, $\epsilon$ do not depend on $x$. Due to the
above symmetry (called shift symmetry from now on), one can
easily derive the corresponding Ward identity and show that
$G_{R\chi}=0$ identically at $G_\chi=0$. Furthermore, from the
Ward identity for the fermion propagator, one can show that
$\mu_R$ will vanish
identically if we choose $K=1/8$ in addition to $G_\chi=0$.
This analytic result tells us that at $G_\chi=0$, $K=1/8$,
there is no mixing between the fermion and mirror-fermion,
and the renormalized Yukawa coupling of the mirror-fermion
is zero. Therefore, the mirror-fermion is massless and is not
coupled to the scalar and fermion fields. In other words,
at $G_\chi=0$ and $K=1/8$, the mirror-fermion behaves like
the massless right-handed neutrino and is decoupled from the
physical world~\cite{LINWIT}.
We call the above choice of the bare parameters
the mirror-fermion decoupling limit.

At the energy scale of the weak interaction (around 100 $GeV$),
all gauge couplings are weak. We therefore take the zeroth-order
approximation by ignoring all gauge fields. We think that the only
possible sources of nonperturbative physics in the electroweak theory
are the Higgs and heavy fermion sectors.
Notice that the shift symmetry of our mirror-fermion model is destroyed
if the gauge fields are put in. Mirror-fermion will be coupled to the
physical world via gauge interactions even if we set $G_\chi=0$,
$K=1/8$. We simply assume that the zeroth-order approximation
reported here serves as a good approximation to the electroweak
theory without worrying how to decouple the mirror-fermion in the
full theory.

\section{Vacuum Stability and the $\beta$-Functions}         \label{s3}

The upper bound on the Higgs mass comes from the notion of
triviality~\cite{LANPOM}.
As for the lower bound, notions of triviality and vacuum stability
are both required~\cite{DUPHSH}.

In order that a quantum field theory is self-consistent, its energy
spectrum has to be bounded from below. Or we say that its vacuum
has to be stable. Since the minimum of the effective potential
corresponds to the ground state (i.e. vacuum) of the theory,
a self-consistent quantum field system should have an effective
potential which is convex. Namely, as the field strength increases,
the effective potential cannot go down to negative values indefinitely.
In terms of the language of renormalization group, the running scalar
coupling constant cannot go to negative values as we increase the
energy scale~\cite{LMMW}.

What describe how the couplings change as the
energy scale changes are the $\beta$-functions.
The exact $\beta$-functions and effective potential are usually unknown.
People rely on loop expansions to derive them. However, loop expansions
can break down when the system becomes strongly interacting, we must
go beyond perturbation theory. At the nonperturbative level, we find
that the partition function of the Euclidean mirror-fermion model simply
blows up when $\lambda$ is negative. If this is so, then the effective
potential is not defined because the theory is not self-consistent.
In order that the theory exists, $\lambda$ has to be positive. Then the
partition function exists and the effective potential can be defined.
Since the exact effective potential is defined via a Legendre
transformation, its shape will be convex and has a minimum once it exists.
Hence, at the nonperturbative level, once $\lambda$ is positive, the
theory will have a stable vacuum.

If the theory is trivial in the continuum limit, then its exact
$\beta$-functions will behave qualitatively like the one-loop ones.
There we can see that at some large $G_\psi$ values,
$\beta$-function of $g$ can turn negative.
This shows that $g$ can become negative
as we go to higher energy scales, thus spoiling the stability of the
vacuum. In order that the vacuum state be stable,
$g_R$ cannot be too small. Hence, we can obtain a
vacuum stability lower bound on $g_R$ as a function of
$G_{R\psi}$. Since the smallest value $\lambda$ can take without destroying
vacuum stability is zero (or $0^+$), the lower bound on $g_R$ naturally
corresponds to a very small but positive $\lambda$~\cite{LMMW}.

 Thus in lattice study of the SU(2) mirror-fermion model, we can first
decouple the unwanted mirror-fermion by setting $G_\chi=0$, $K=1/8$. If the
mirror-fermion model is trivial in the continuum limit, Monte Carlo data
on $g_R$ and $G_{R\psi}$ obtained at $\lambda=\infty$ and $\lambda=0^+$
will give us upper and lower bounds on the Higgs mass as a function of
the heavy fermion mass respectively.
 It goes without saying that both bounds are cutoff dependent.

We also carry out one-loop renormalized perturbation theory on
the lattice and calculate $\beta$-functions accordingly. We therefore
can estimate the finite size effects and finite cutoff effects.

\section{The Simulation and Data}                       \label{s4}

The phase structure of the model was first explored. (See~\cite{LIMOWI}
for details.) We find that the phase boundary
between the symmetric (PM) phase and the broken symmetry (FM)
phase is a second-order
phase transition line where the cutoff can be removed.

Numerical simulations were performed on $6^3 \cdot 12$ and $8^3\cdot 16$
lattices at $\lambda=10^{-6}$ and $\infty$.
The small positive value of $\lambda$ was chosen to mimic $\lambda=0^+$.
The Yukawa coupling $G_\chi$ was kept at zero, the fermion hopping
parameter $K$ was kept near $1/8$ on the finite lattice for exact
decoupling of the mirror doublets~\cite{FLMMPTW,LINWIT}.
Since it is the FM phase that is physically relevant,
$\kappa$ and $G_\psi$ are tuned such that we approach criticality from
within the FM phase and
have a reasonable cutoff to reduce finite size effects.
Since Ward identities give us some analytic results for renormalized
quantities at $G_\chi=0$, we put those results into our Fortran program
to ``guide" our Monte Carlo simulations. This helps to reduce the
fluctuations.

\begin{figure}
\vspace*{5.0cm}
\leftline{\hfill\vbox{\hrule width 0cm height0.001pt}\hfill}
\vspace*{1.4truein}		
\leftline{\hfill\vbox{\hrule width 0cm height0.001pt}\hfill}
\fcaption{Higgs mass bounds as a function of heavy fermion mass. Curves are
from one-loop perturbative calculations in the continuum
at $\xi=1.0$ and $1.2$.}
\label{figure1}
\end{figure}

\newpage
\section{Conclusions}                                     \label{s5}

Our numerical data are presented in Fig. 1.
At all points of the simulation, we find that all fermion doublers
are decoupled as expected.

It is seen that our Monte Carlo data are in qualitative
agreement with one-loop perturbative results. This may not be
so surprising since tree unitarity (which requires probability
conservation) upper limit on $G_{R\psi}^2$ is $2\pi$. Below this
value, the system should be weakly interacting. So far, no evidence
for an ultra-violet stable fixed point has been found.
Since all qualitative features of the one-loop $\beta$-functions are
supported, our mirror-fermion
models are likely to be trivial in the infinite cutoff limit.
Assuming that the SU(2) mirror-fermion model is trivial in the
continuum limit, we obtain, in zeroth-order approximation,
upper and lower bounds on the Standard Model Higgs mass as a function
of the heavy fermion mass. If the heavy fermion in our model
is top quark with its mass
around $180\, GeV$, then we claim at $\xi = 1.0$ that Standard Model
Higgs has its mass between $100$ and $800\, GeV$.

Although the observed qualitative behaviour is certainly consistent
with the one-loop perturbative scenario implying the triviality
of the continuum limit, one has to keep in mind that present
simulations are done at relatively low cut-offs.
In particular, the evolution of the couplings towards smaller
values at decreasing cut-offs should be investigated in the
future in order to make sure that the model is indeed trivial.



\section{References}

\begin{thebibliography}{9}
%
\bibitem{WS}
K. G. Wilson, {\it New Phenomena in Sub-nuclear Physics} (Plenum,
New York, 1977);\\
J. B. Kogut, L. Susskind, {\it Phys. Rev.} {\bf D11} (1975) 395.
%
\bibitem{SS}
J. Smit, {\it Acta Phys. Polon.} {\bf 17} (1986) 531;\\
P. V. D. Swift, {\it Phys. Lett.} {\bf 145B} (1984) 256.
%
\bibitem{IM}
I. Montvay, {\it Phys. Lett.} {\bf 199B} (1987) 89.
%
\bibitem{LMMW}
L.\ Lin, I.\ Montvay, G.\ M\"unster, H.\ Wittig,
{\it Nucl.\ Phys.}\ {\bf B355} (1991) 511.
%
\bibitem{LINWIT}
M.F.L.\ Golterman, D.N.\ Petcher,
{\it Phys.\ Lett.}\ {\bf B225} (1989) 159;\\
L.\ Lin, H.\ Wittig,
{\it Z.\ Phys.}\ {\bf C54} (1992) 331.
%
\bibitem{LANPOM}
L.D.\ Landau, I.Y.\ Pomeranchuk,
{\it Dokl.\ Akad.\ Nauk SSSR} {\bf 102} (1955) 489.
%
\bibitem{DUPHSH}
M.J.\ Duncan, R.\ Philippe, M.\ Sher,
{\it Phys.\ Lett.}\ {\bf B153} (1985) 165; \\
M.\ Sher,
{\it Phys.\ Rep.}\ {\bf 179} (1989) 273.
%
\bibitem{LIMOWI}
L.\ Lin, I.\ Montvay, H.\ Wittig,
{\it Phys.\ Lett.}\ {\bf B264} (1991) 407.
%
\bibitem{FLMMPTW}
C.\ Frick, L.\ Lin, I.\ Montvay, G.\ M\"unster, M.\ Plagge,
T.\ Trappenberg, H.\ Wittig,
{\it Nucl.\ Phys.}\ {\bf B397} (1993) 431;
{\it Nucl.\ Phys.}\ {\bf B}({\it Proc.\ Suppl.}){\bf 30} (1993) 647.
%
\end{thebibliography}
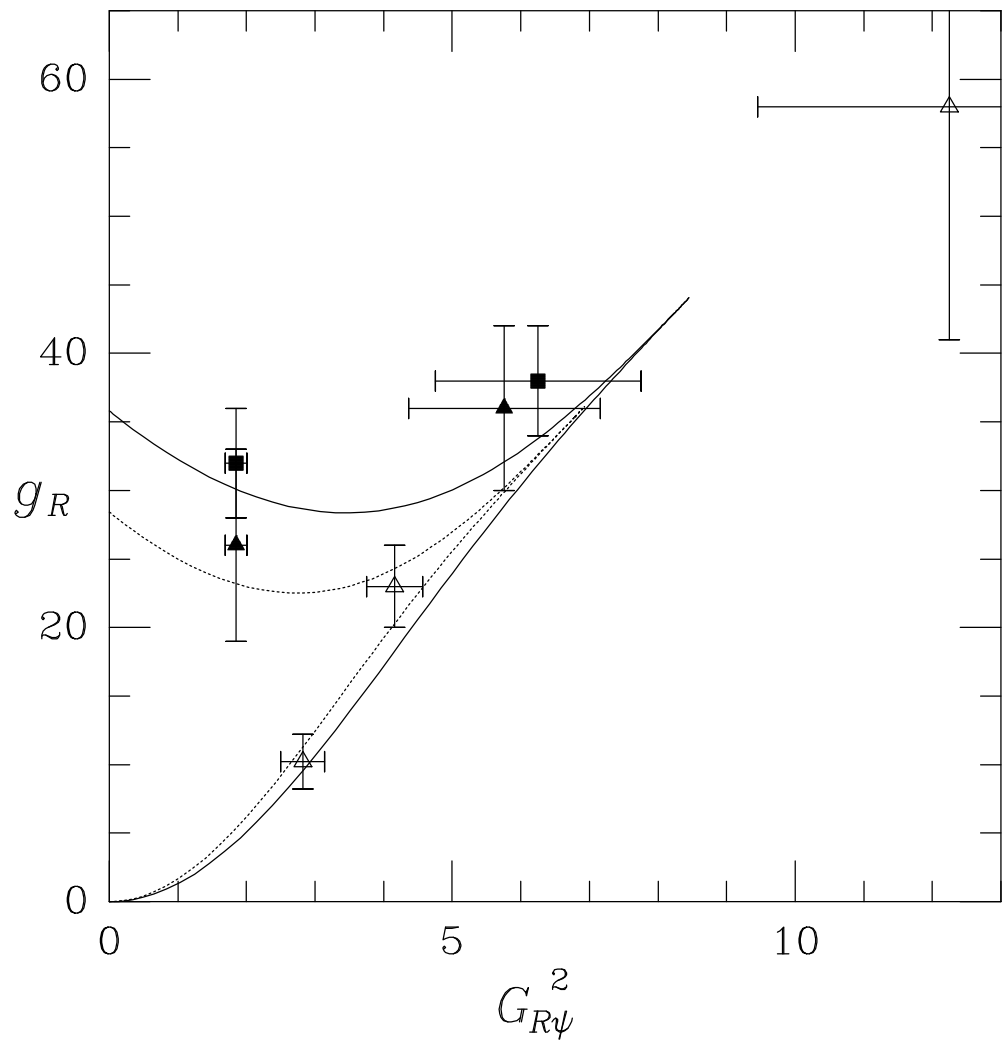
\end{document}